\documentclass[fleqn,usenatbib]{mnras}

\usepackage[T1]{fontenc}

\DeclareRobustCommand{\VAN}[3]{#2}
\let\VANthebibliography\thebibliography
\def\thebibliography{\DeclareRobustCommand{\VAN}[3]{##3}\VANthebibliography}

\usepackage{graphicx}	
\usepackage{amsmath}	
\usepackage{amssymb}	
\usepackage{comment}    
\usepackage{rotating}   
\usepackage{hyperref}   
\usepackage{caption}
\usepackage{subcaption}
\usepackage{soul}
\usepackage[normalem]{ulem}
\usepackage{dirtytalk}
\usepackage[dvipsnames]{xcolor}

\title[Magnetic field measurement in TMC-1C]{Magnetic field measurement in TMC-1C using 22.3 GHz CCS Zeeman splitting}

\author[Koley et al.]{
Atanu Koley$^{1,2}$ \thanks{E-mail: atanuphysics15@gmail.com}, Nirupam Roy$^{2}$, Emmanuel Momjian$^{3}$, Anuj P. Sarma$^{4}$, Abhirup Datta$^{5}$
\\
$^{1}$Joint Astronomy Programme, Indian Institute of Science, Bangalore 560012, India\\
$^{2}$Department of Physics, Indian Institute of Science, Bangalore 560012, India\\
$^{3}$National Radio Astronomy Observatory, P. O. Box O, 1003 Lopezville Road, Socorro, NM 87801, USA\\
$^{4}$Physics and Astrophysics Department, DePaul University, 2219 N. Kenmore Ave., Chicago, IL 60614, USA\\
$^{5}$Discipline of Astronomy, Astrophysics and Space Engineering, Indian Institute of Technology Indore, Indore 453552, India
}

\date{Accepted XXX. Received YYY; in original form ZZZ}

\pubyear{2022}

\begin{document}
\label{firstpage}
\pagerange{\pageref{firstpage}--\pageref{lastpage}}
\maketitle

\begin{abstract}

Measurement of magnetic fields in dense molecular clouds is essential for understanding the fragmentation process prior to star formation. Radio interferometric observations of 
CCS 22.3 GHz emission, from the starless core TMC-1C, 
have been carried out with the Karl G. Jansky Very Large Array to search for Zeeman splitting of the line in order to constrain the magnetic field strength. Toward a region offset from the dust peak, we report a detection of the Zeeman splitting of the CCS 2$_{1}$ - 1$_{0}$ transition, with an inferred magnetic field of $\sim$2~mG. If we interpret the dust peak to be the core of TMC-1C, and the region where we have made a detection of the magnetic field to be the envelope, then our observed value for the magnetic field is consistent with a  subcritical mass-to-flux ratio envelope around a core with supercritical mass-to-flux ratio.  The ambipolar diffusion timescale for the formation of the core is consistent with the relevant timescale based on chemical modeling of the TMC-1C core. This work demonstrates the potential of deep CCS observation to carry out future measurements of magnetic field strengths in dense molecular clouds and, in turn, understand the role of the magnetic field in star formation.

\end{abstract}




\section{Introduction}
\label{Introduction}
Starless cores are an ideal place to study star formation in molecular clouds. Here, due to the absence of existing stars, no feedback effect influences the star formation process. Starless cores are broadly divided into two categories: evolved and young starless cores \citep{2004A&A...416..191T,2011ApJ...739L...4R,2021A&A...653A..15N}. Young starless cores are in the preliminary stages of their evolution and do not have a sufficiently dense central nucleus, whereas evolved cores are believed to be out of the equilibrium stage and in the process of forming stars.

It is well known that stars are formed from dense molecular clouds through gravitational collapse, although the details of the complex processes leading to star formation remain unclear. Although there are several theories on cloud fragmentation and star formation, the debate is still ongoing for whether magnetic field or turbulence is the most dominant factor in the star formation process  \citep{1956MNRAS.116..503M,1999ApJ...526..279P,2001A&A...371..274U,2019FrASS...6...66C,2021A&A...649L..13D,2021Galax...9...41L,2021A&A...652A..69M}. To get a better understanding of the process of star formation, it is crucial to measure the magnetic field strength and its morphology in different density regimes in star-forming prestellar cores. In molecular clouds, Zeeman splitting and dust polarization are the two main methods for studying the magnetic field, with the former providing the line of sight magnetic field strength using spectral lines. One of the spectral lines suitable for magnetic field measurements via Zeeman splitting is the 2$_{1}$ - 1$_{0}$ 22.3 GHz transition of the CCS molecule that has a relatively high Zeeman splitting factor (0.767 Hz $\mu$G$^{-1}$). Furthermore, the critical density of CCS, $\sim$ 10$^{4}$ cm$^{-3}$, is well matched to the typical densities of prestellar dense cores ($\sim$ 10$^{4}$ $-$ 10$^{5}$ cm$^{-3}$; \citealt{2019PASJ...71..117N}). This means that CCS should be observable in these prestellar cores if it is abundant and therefore it is an obvious choice to measure the magnetic field in such prestellar dense cores. Apart from Zeeman splitting measurements, CCS is used to probe density as well as the evolutionary age of dense cores. 

The TMC-1C core has been studied extensively using various molecular transitions (\citealt{2016A&A...594A.117P,2016A&A...586A.110S}). There are existing single dish Zeeman splitting measurements of similar cores TMC-1 and L1521E at 45.4 GHz (yielding magnetic field strength of $\sim$117  $\pm$ 21 $\mu$G in TMC-1 and $\sim$160 $\pm$ 42 $\mu$G in L1521E; \citealt{2019PASJ...71..117N,1999sf99.proc..175S}). In this Letter, we present interferometric observations of CCS toward the TMC-1C core with the aim of measuring the Zeeman effect. The kinematical properties of TMC-1C, along with that of TMC-1 and L1544, based on the observations of CCS and NH$_{3}$, will be reported in a forthcoming paper (Koley, A., 2022, in prep.). We decided to target TMC-1C based on earlier interferometric observations \citep{2011ApJ...739L...4R}, indicating that the CCS 22.3 GHz line intensity and narrow line profile are suitable for Zeeman splitting measurements. 

\begin{figure}
\centering 
\includegraphics[width=3.5in,height=3.3in,angle=0]{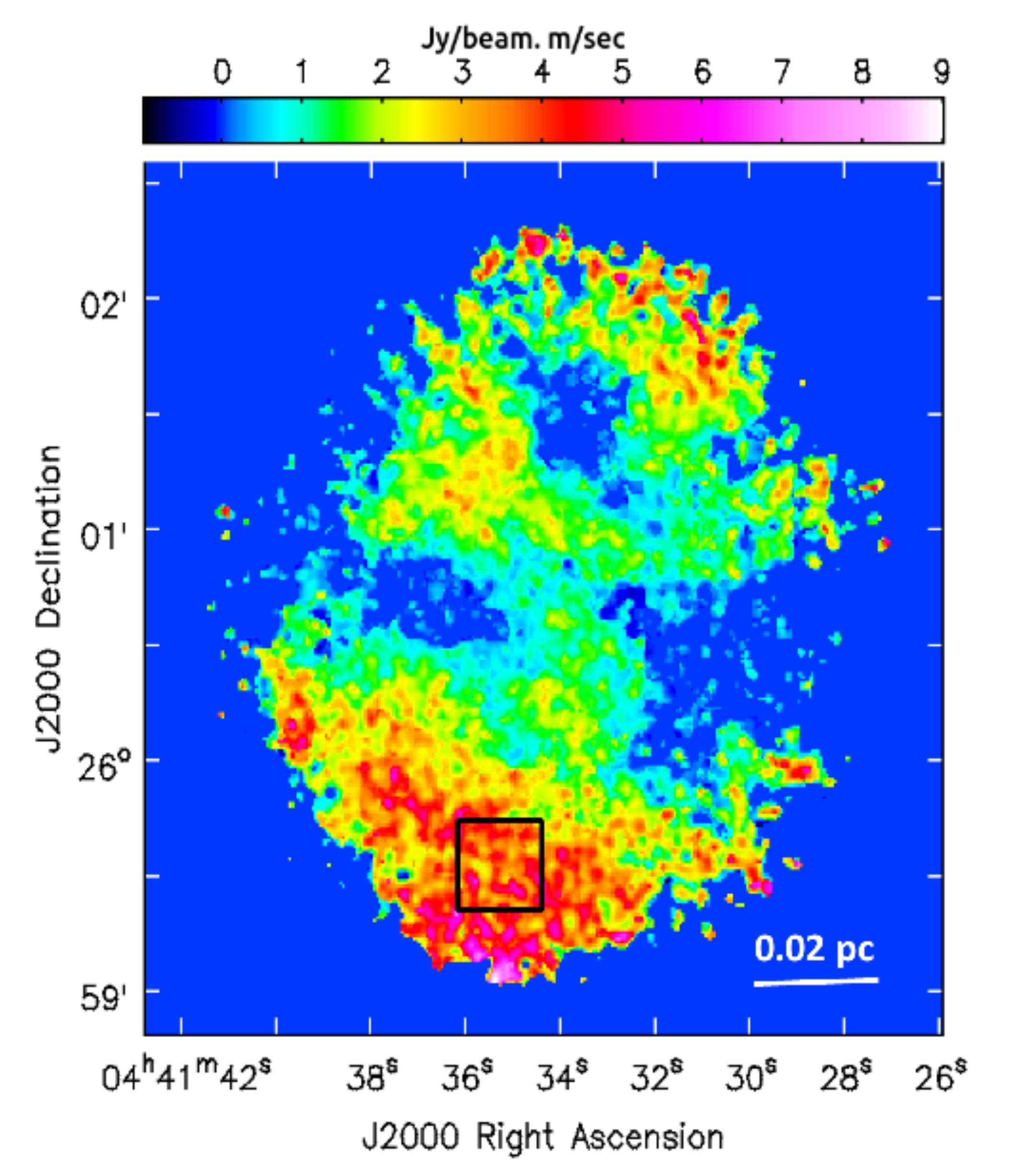}
  \caption{Integrated CCS 22.3 GHz line intensity image (primary beam corrected) of TMC-1C. The black square (from $\alpha = 04^{\rm h}41^{\rm m}36\fs11$, $\delta = 25{\degr}59{\arcmin}22\farcs58$ to   $\alpha = 04^{\rm h}41^{\rm m}34\fs44$, $\delta = 25{\degr}59{\arcmin}43\farcs74$) denotes the region in which we report the detection of the magnetic field due to the Zeeman effect.}
  \label{fig:fig1}
\end{figure}

This Letter is organized in the following way. Summary of the observations and data reduction are provided in section \S\ref{observation and data analysis}. The results on the Zeeman splitting measurement and magnetic field estimation are presented in \S\ref{Results}, and discussed in \S\ref{Discussion}. Finally, we present the conclusions in section \S\ref{Conclusions}.  

\begin{figure*}
\centering 
 \includegraphics[width=3.7in,height=2.8in,angle=0]{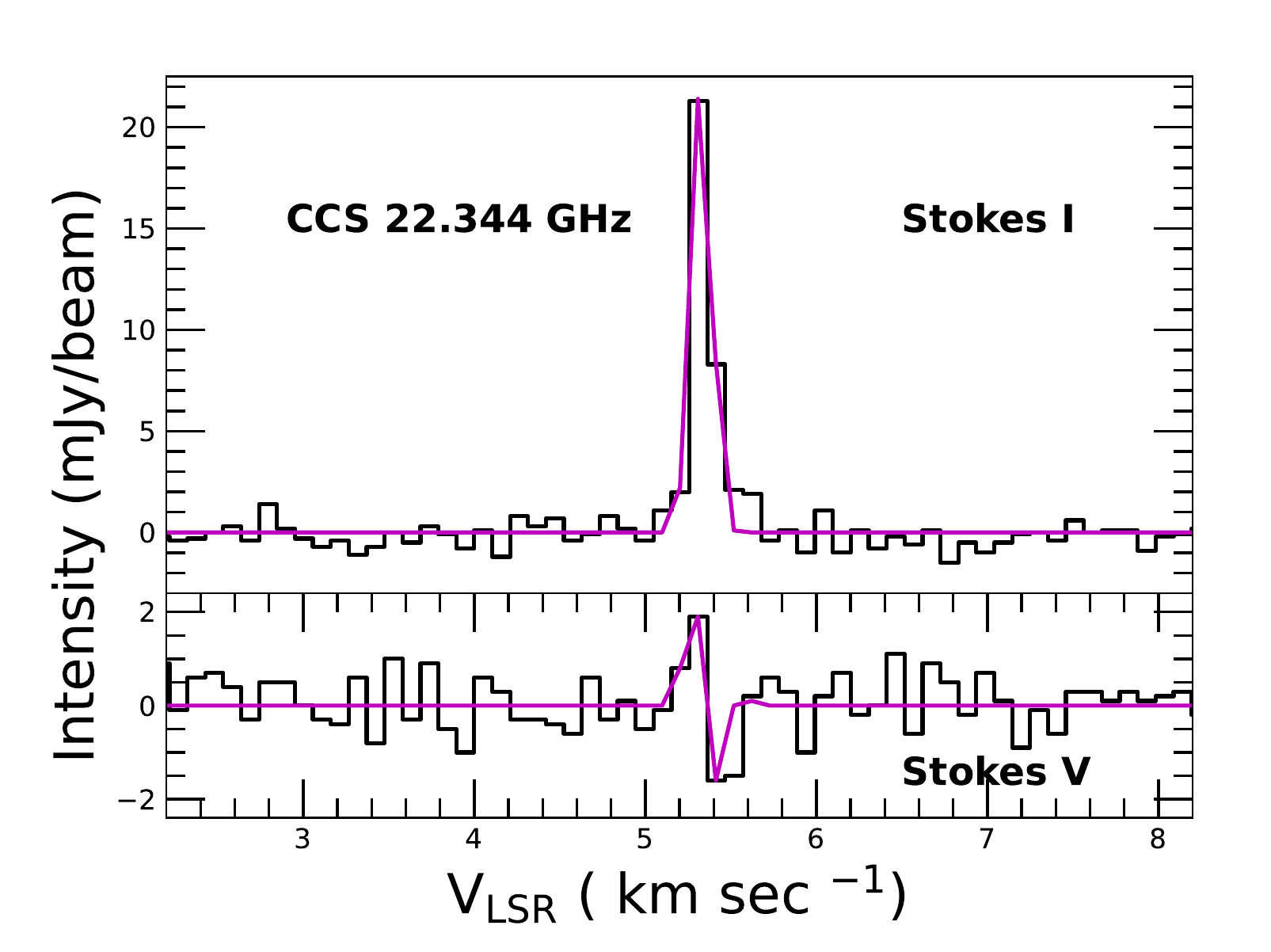}~~ \includegraphics[width=3.7in,height=2.8in,angle=0]{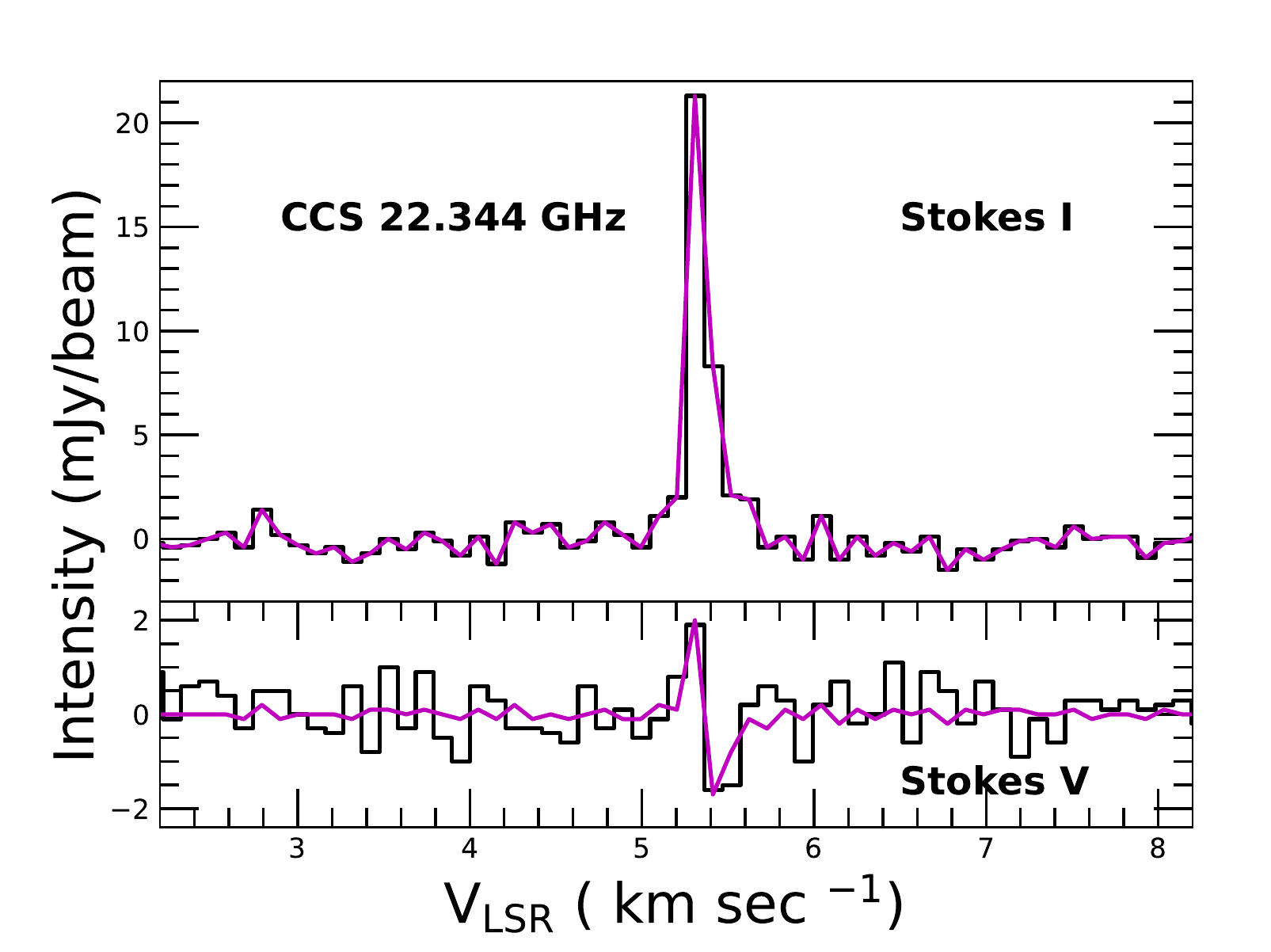}
  \caption{Left: Average Stokes $I$ (upper panel---black histogram-like line) and Stokes~$V$ (lower panel---black histogram-like line) profiles of CCS 22.3 GHz line toward the black square region in Figure~\ref{fig:fig1}. The magenta curve in the upper panel shows the Gaussian component that we fitted to the Stokes~$I$ profile. In the lower panel, the magenta  solid line is the fit to the Stokes~$V$ profile for the estimated $B_{\text{los}}$ = $-1.80 \pm 0.38$ mG. Right: Average Stokes~$I$ (upper panel---black histogram-like line) and Stokes~$V$ (lower panel---black histogram-like line) profiles toward the same region marked by the black square. The solid magenta  line in the upper panel is the observed data (that is, the same as the black histogram-like line, but plotted as a continuous curve). The magenta  solid line in the lower panel is the fit to the Stokes~$V$ profile for the estimated $B_{\text{los}}$ = $-$2.40 $\pm$ 0.44 mG from the one-sided derivative of Stokes~$I$.}
   \label{fig:fig2}
\end{figure*}

\section{Observations and Data Analysis}
\label{observation and data analysis}
The data for this work were taken using the Karl G. Jansky Very Large Array (VLA) as part of the proposal VLA/13A-112. The observations were carried out in 2013 February and March (over 13 observing sessions). More details of the observations are provided in Appendix~\ref{appendix1}. All the initial flagging and calibration of the data were performed both in the scripted CASA (Common Astronomy Software Applications package) pipeline of the National Radio Astronomy Observatory (NRAO)\footnote{The National Radio Astronomy Observatory is a facility of the National Science Foundation operated under cooperative agreement by Associated Universities, Inc.} and manually in AIPS (Astronomical Image Processing System; version 31DEC19). This was done as a consistency check of the analysis. There was no noticeable difference found between the output of the two methods. For imaging, we used the CASA task \textbf{tclean} with the multiscale deconvolver to pick up all the different structures in the core in total intensity. The scales that we used correspond to $3\farcs2$, $7\farcs2$, $16\farcs0$, $24\farcs0$, and $32\farcs0$. We note that multiscale CLEANing produces a single image that is restored with a synthesised beam that corresponds to the full angular resolution of the data. Due to the low signal to noise, we did not CLEAN the Stokes~$V$ image cube.

\section{Results} 
\label{Results}
Magnetic field values in the interstellar medium are relatively low ($\sim$ $\mu$G - mG). Thus, the expected Stokes~$V$ signal for typical magnetic field values in a dense core is very weak, requiring deep observations with high spectral resolution to detect the Zeeman splitting signal. In astrophysical environments, spectral lines are significantly broadened by thermal and non-thermal motions. Therefore, the Zeeman splitting is much smaller than the line width, which in turn results in the measurement of only the line-of-sight component ($B_{\rm los}= B{\rm cos}\theta$) of the magnetic field \citep{crutcher93,sault90}. In this small splitting approximation, the fitting formulae for Stokes~$V$ with Stokes~$I$ is:
\begin{eqnarray}\label{equation1}
V = aI + \frac{b}{2}\frac{dI}{d\nu}
\end{eqnarray}
where $a$ is the gain difference between the left circular polarization (LL) and the right circular polarization (RR), $b = zB\mathrm{cos}\theta$, and $z$ is the Zeeman splitting factor \citep{greisen17a}.  We used \textbf{XGAUS} and \textbf{ZEMAN} in AIPS to fit the observed Stokes~$I$ and $V$ profiles. 

The integrated intensity map of the CCS 22.3 GHz emission in TMC-1C is shown in Figure~\ref{fig:fig1}. This figure reveals a  very patchy appearance spread over the field of view. Figure~\ref{fig:fig2} shows the Stokes~$I$ and $V$ profiles toward the region highlighted by the black box in Figure~\ref{fig:fig1}. The Stokes~$V$ profile (bottom panels of Figure~\ref{fig:fig2}) has an S-shape that usually suggests a Zeeman splitting detection. Typically, at least 6 channels across the spectral line is considered to be necessary in order to claim a Zeeman splitting detection. Since the profile in Figure~\ref{fig:fig2} marginally spans 6 channels ``null-to-null'', we claim that the observations reported in this Letter represent a detection. To eliminate the possibility of any systematic effects that may lead to a false S-curve, we checked each individual session's data, and found that the Stokes~$V$ profiles were indeed noise-like with no spurious features that may be attributed to instrumental effects. We obtained the magnetic field value in two different ways: (1) We fitted the observed Stokes~$I$ profile with a single Gaussian component, then found the derivative of that component and fitted it to Stokes~$V$ using equation\,(\ref{equation1}) to derive $B_{\text{los}} = -1.80 \pm 0.38$ mG, and (2) We calculated the one-sided derivative of the observed Stokes~$I$ profile and fitted it to Stokes~$V$, again using equation\,(\ref{equation1}) to find $B_{\text{los}}  = -2.44 \pm 0.44$ mG. Within the errors, both methods agree on the magnetic field value, which we adopt to be 2 mG for further analysis in section \S\ref{Discussion}. 

We note that, for single dish observations, the beam squint can produce false Zeeman signal in presence of velocity gradient of the line emission. For interferometric observations, the prominent effect is leakage of Stokes I into Stokes V \citep{sault90}. In our modeling, leakage is included in Eqn.~\ref{equation1} to account for this. Further, our detected Stokes V signal indicates a magnetic field strength of $\sim 2$ mG. As the observed line profile is fairly narrow, we can clearly rule out, from our observations, any large velocity gradient that will be required, along with beam squint, to produce a Zeeman-like signal.

\section{Discussion} 
\label{Discussion}
This work presents the interferometric study of CCS 22.3 GHz emission from the TMC-1C region. The emission is found to be widespread and patchy, while the line profile is very narrow and consistent with earlier shallow observations \citep{2011ApJ...739L...4R}. For one region within the primary beam of the VLA, we also have a detection of the magnetic field, based on the Zeeman splitting of the transition. If the signal is real, the determined magnetic field is about 2 mG. 

\subsection{CCS Emission}
\citet{1992ApJ...392..551S} measured the column density of CCS in TMC-1C with the Nobeyama 45 meter telescope and found N(CCS) $\sim$ 3.1 $\times$ 10$^{13}$ cm$^{-2}$. As the diffuse emission is resolved out, only 33\% of this column density is detected in our VLA observations. In Figure~\ref{fig:fig3}, the integrated intensity of CCS 22.3 GHz line is superimposed on the 1200 $\mu$m dust emission from a larger region \citep{2008A&A...487..993K}. The white square denotes the area in which we have detected the magnetic field. A clear indication of a spatial offset between the CCS and dust emission is observed as seen in Figure~\ref{fig:fig3}. The CCS emission peaks are shifted from the dust emission peaks, and there is very little CCS emission from the region where the dust emission peaks within the VLA's primary beam. CCS is a carbon-bearing species, and it is depleted during the evolution of the core. Thus, the depletion of CCS is observed toward the extremely dense nucleus where dust or nitrogen-bearing species are present in copious amounts. \citet{tatematsu2014} and \citet{seo2019} have observed a similar offset in the CCS emission from Orion A and L1495-B218, respectively.

We note that there is some variation of the spectral shape of the CCS emission across the region, with a hint of two components in the Stokes~$I$ profile in some regions (including the region where Zeeman splitting is detected). The broader component is relatively weak, similar to what is observed for NH$_3$ in L1688 \citep{2020A&A...640L...6C}. The reduced chi-square for a two-component Gaussian fit to the Stokes $I$ spectra is slightly better than that of a single component fit, and indicates that the weak, broad component is likely to be real. With very similar central velocities, these are likely to be associated components, similar to two NH$_3$ components in L1688 (\citealt{2020A&A...640L...6C}, also see \citealt{kama2015}). However, the spectral resolution and the sensitivity of the current observations are not adequate to reliably fit the Stokes~$V$ signal with two components; hence, we have used the one-sided derivative and single component fit methods to determine an effective single average value for the magnetic field.

\subsection{Magnetic Field}
In this section we discuss the role of the magnetic field during the evolution of a prestellar core, which can be determined by the ratio of the observed to the critical mass-to-flux ratios, given by the equation:
\begin{equation}
{\lambda_c \equiv \frac{(M/\Phi)_{\rm obs}}{(M/\Phi)_{\rm critical}} = 7.6 \times 10^{-21} {\frac{N(\text{H}_{2})}{B_{\text{total}}}}}
\end{equation}
where N(H$_{2}$) is in cm$^{-2}$, $B_{\rm total}$ is in $\mu$G and $({M}/{\Phi})_{\rm critical}= {1}/{\sqrt{4\pi^{2} G}}$ \citep{1978PASJ...30..671N}. Earlier, \citet{caselli08} have reported  N(H$_{2}$) to be $8.5\times10^{22}$ cm$^{-2}$ towards the dust emission peak in TMC-1C. With a factor of 2 lower dust emission (compared to the dust emission peak) toward the region enclosed by the white box in Figure \ref{fig:fig3}, these observations, therefore, imply N(H$_2$) $\sim 4.25\times10^{22}$ cm$^{-2}$ for the region in which we have reported a detection of the magnetic field. This is consistent with a number density of (1.4 $\pm$ 0.8) $\times$ 10$^{5}$ cm$^{-3}$ \citep{2021A&A...653A..15N} and linear size of $\sim$ 0.1 pc \citep{1998ApJ...504..207B} for this region. Based on their chemical modeling, \citet{1992ApJ...392..551S} predicted a fractional abundance ${n_{\rm CCS}}/{n_{\rm H_{2}}} =(2 - 6) \times 10^{-10}$  i.e., N(H$_2$) $\sim (5 - 15) \times 10^{22}$~cm$^{-2}$ from the Nobeyama observations. This was for the position coinciding with the pointing centre of our VLA observations, and with a factor of 1.5 lower dust emission (compared to the VLA pointing centre) toward the region of our interest, N(H$_2$) is expected to be $\sim (3.3 - 10) \times 10^{22}$~cm$^{-2}$. We note that the value of N(H$_2$) can also be estimated from the average 1200$\mu$m dust emission intensity, which is $\sim$ 18 mJy beam$^{-1}$ in the region marked by the white box in Figure~\ref{fig:fig3}, and the conversion factor reported in \citet{2008A&A...487..993K}. This gives an N(H$_2$) $\sim 1.2 \times 10^{22}$~cm$^{-2}$, which is significantly lower than the above values and results in an unusually small value of $\lambda_c$. Hence, for our analysis below, we adopt N(H$_2$) $\sim (3.3 - 10) \times 10^{22}$~cm$^{-2}$.

Our observations of TMC-1C resulted in a detection of the Zeeman effect with a line of sight magnetic field value of $\sim2$ mG. As the CCS emission is weak at the position of the dust peak, and the Stokes~$V$ signal is detected only in a region offset from that peak, the inferred magnetic field is possibly associated with the surrounding envelope of the dense core, where the magnetic field is expected to be dynamically more important. 
We note that magnetic field of the order of a few mG for the density regime probed by CCS is broadly consistent with earlier observations (e.g. \citealt{2022ApJ...925...30L} and references therein). To assess the implications of this measurement, we calculate the mass-to-flux ratio by considering the line-of-sight magnetic field is approximately equal to the total magnetic field ($B_{\rm total} = B_{\rm los}$). For N(H$_{2}$) $= (3.3 - 10)\times10^{22}$ cm$^{-2}$, this gives $\lambda_c = 0.13 - 0.38$, suggesting a subcritical envelope. This is consistent with the general picture of  supercritical starless accreting core, surrounded by a subcritical envelope. Indeed,  earlier  large scale  observations  of the  Taurus  Molecular  Cloud  have  shown  the  cloud  to be subcritical \citep{2008ApJ...680..420H, 2012MNRAS.420.1562H}. A logical question to ask is whether the timescale of formation and growth for the starless accreting core in TMC-1C is consistent with other observations.

If during the stage of evolution and gravitational collapse, a supercritical core is formed via the ambipolar diffusion process, we can make an order of magnitude estimation of the ambipolar diffusion timescale $t_{\rm AD}$ \citep{2004fost.book.....S,2009RMxAC..36..278K, 2014MNRAS.443..230H,2019FrASS...6....5H}:  
\begin{equation}
t_{\rm AD} \approx 3 \times 10^{6} \left(\frac{n_{{\rm H}_{2}}}{10^{4 }~{\rm cm}^{-3}}\right)^{3/2}\left(\frac{B}{30~\mu {\rm G}}\right)^{-2}\left(\frac{L}{0.1~{\rm pc}}\right)^{2} \,{\rm yr}
\end{equation}
where $n_{{\rm H}_{2}}$ is the number density of the hydrogen molecule, $B$ is the magnetic field and $L$ is the length scale over which the ambipolar diffusion occurs in a system. 
For coherent dense cores with subsonic non-thermal motions \citep{2010ApJ...721..686P}, the relevant scale is the size of the core \citep{hennebelle2019, 1991ApJ...371..296M}. For TMC-1C, the size of the core is $\approx$ 0.1 pc \citep{1998ApJ...504..207B, 1998ApJ...504..223G}. The estimated number density of TMC-1C, from analysis based on multiple observed transitions of CS, is as high as 1.4 $\pm$ 0.8 $\times$ 10$^{5}$ cm$^{-3}$ \citep{2021A&A...653A..15N}. Hence, if we adopt $L$ = 0.1 pc, $n_{{\rm H}_{2}}$ = 1.5 $\times$ 10$^{5}$ cm$^{-3}$ and take $B$ $\approx$ 2 mG ($ B_{\rm los}$), the ambipolar diffusion will make the core significantly supercritical within a timescale of $\sim$ 4 $\times$  $10^{4}$~yr. We note that the diffusion time scale may further decrease by a factor of few if the ambient turbulent flow is comparable to or greater than the Alfven velocity \citep{2014ApJ...794..127K}. This may look like a rather short timescale compared to the age of TMC-1C, estimated to be $>$ 3 $\times$ 10$^{5}$ yr \citep[based on N$_{2}$H$^{+}$ abundance]{2007ApJ...671.1839S} to $\approx$ 1 Myr \citep[based on detailed astrochemical models;][]{2021A&A...653A..15N}. However, \citet{2007ApJ...671.1839S} have also concluded that the TMC-1C core is much younger or has undergone significant mass infall from the surrounding  envelope with a corresponding timescale of $\sim$ 10$^{4}$ yr. Thus, the relevant timescale matches well with the ambipolar diffusion timescale.

\section{Conclusions} 
\label{Conclusions}
We have observed the 22.3 GHz CCS line toward the starless core TMC-1C with the VLA. We report in this Letter the  detection of CCS Zeeman splitting, the first of its kind, with a radio interferometer. Our inferred line of sight magnetic field toward TMC-1C is $\sim$2 mG. Future observations with higher sensitivity and higher spectral resolution will be necessary to confirm these results. The Zeeman detection is toward a region to the southwest of the 1200~$\mu$m dust peak. If we interpret the dust peak to be the core of TMC-1C, then our observed magnetic field may be associated with the envelope region around the core. This is consistent with the general expectation of a supercritical core and a subcritical envelope in an accreting starless core like TMC-1C where the magnetic field in the envelope may be dynamically important, and the accreting core grows via the ambipolar diffusion process. The ambipolar diffusion timescale of core formation is found to be in agreement with the mass infall timescale for TMC-1C reported by previous authors. The detection reported here is a crucial step highlighting the possibility of future interferometric measurements of magnetic fields in dense molecular clouds using CCS transitions. 

\section*{Acknowledgements}
We thank the reviewer for many useful comments and valuable suggestions that have helped in improving this paper significantly. We are grateful to Thushara Pillai for useful discussions and valuable inputs. A.K. would like to thank DST-INSPIRE (IF160553) for a fellowship.

\section*{Data Availability}

All data used in this study are available publicly from the VLA online archive. All reduced data will be shared at reasonable request to the corresponding author.



\bibliographystyle{mnras}
\bibliography{mnras} 


\appendix

\section{Details of observations}
\label{appendix1}
Details of the VLA observations are summarized in table~\ref{tab:table1}.

\vspace{80mm}
\begin{table}
	\centering
\caption{Summary of observational parameters.}
	\label{tab:table1}
	\begin{tabular}{lc} 
\hline
VLA proposal ID & 13A-112 \\
Date of Observations & 2013 February 7--March 29 \\
Configuration & D  \\
Observing Band & K (18.0-26.5~GHz) \\	
R.A. of field center (J2000) & $04{^{\rm h}} 41{^{\rm m}} 34\fs316$ \\
Dec. of field center (J2000) & $+26{^\circ} 00^{\prime} 42\farcs592$ \\
Calibrators & 3C~ 147 \\
~ & J0440$+$2728\\
Spectral line: & \\
Rest Frequency & 22.344~GHz \\
Observing Bandwidth & 
4~MHz \\
Number of channels & 
512 \\
Channel spacing & 0.1 km~s$^{-1}$ \\
Primary beam (HPBW) & ${\sim 135^{\prime\prime}}$ \\
Synthesised beam (FWHM) & $3\farcs57 \times 3\farcs40$\\
On-source time & ${\sim 13}$~hr~$30$~min \\
RMS noise (per channel)  &${\sim 0.8}$~mJy~beam$^{-1}$ \\
\hline
	\end{tabular}
\end{table}


\vspace{-80mm}
\section{Comparison of CCS and dust emission}
\label{appendix2}
\begin{figure}
  \includegraphics[width=3.0in,angle=0]{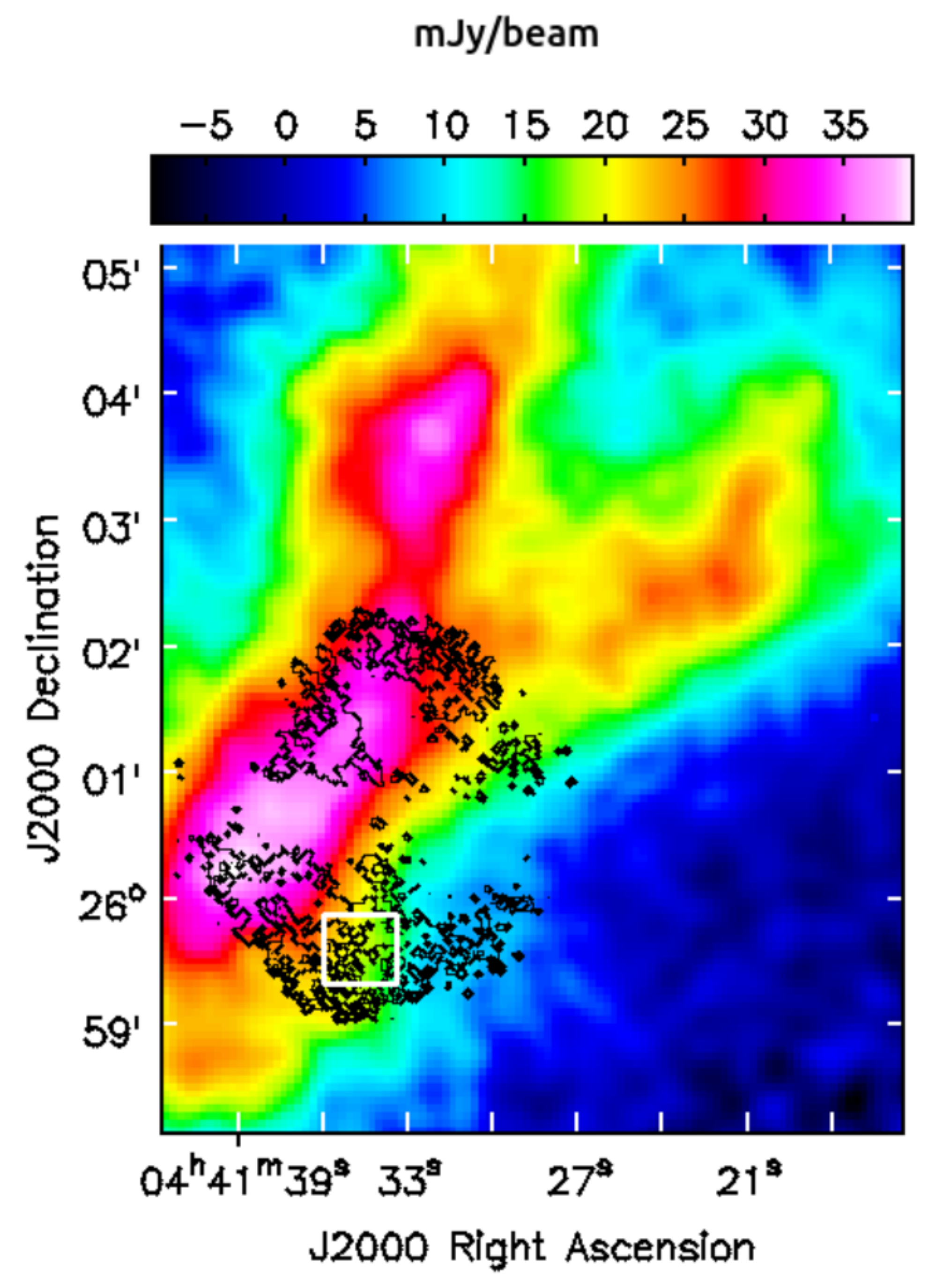}
  \caption{Overplot of the integrated intensities of CCS 22.3 GHz emission and the 1200 $\mu$m dust emission \citep{2008A&A...487..993K} at a resolution of $10\farcs80 \times 10\farcs50$. Dust integrated intensity is shown in the color plot and CCS emission is in the contours. The contour levels are at 1.76, 3.58, 5.40, and 7.22 Jy beam$^{-1}$. The white square denotes the region where we report the detection of the magnetic field.}
 \label{fig:fig3}
\end{figure}

The archival $1200~\mu$m public data for TMC-1C region\footnote{Data available through the Strasbourg astronomical Data Centre (CDS): \url{https://vizier.u-strasbg.fr/viz-bin/VizieR-3?-source=J/A\%2bA/487/993/object}} is used to compare the dust and CCS emission distribution in this region. The dust map was acquired using the Max-Planck Millimeter Bolometer (MAMBO) array at the IRAM 30-meter telescope \citep{2008A&A...487..993K} . Figure~\ref{fig:fig3} is an overlay of the CCS 22.3 GHz integrated intensity with the MAMBO $1200~\mu$m emission, that shows the offset of the dust peak with respect to the prominent CCS emission. The region indicated by white square is inferred to be in the envelope around the dense core where the Zeeman splitting signal is detected. 

\bsp	
\label{lastpage}

\end{document}